\documentclass[11pt]{article}
\usepackage{amsmath,amssymb}
\usepackage{graphicx,color}
\usepackage{cite}

\DeclareMathOperator{\re}{Re}
\DeclareMathOperator{\im}{Im}
\DeclareMathOperator{\Tr}{Tr}

\def\({\left(}
\def\){\right)}

\newcommand{\de}{\partial}
\newcommand{\be}{\begin{equation}}
\newcommand{\ba}{\begin{eqnarray}}
\newcommand{\ea}{\end{eqnarray}}
\newcommand{\ee}{\end{equation}}

\newcommand{\we}{\wedge}

\newcommand{\f}{\frac}
\newcommand{\s}{\sqrt}
\newcommand{\vp}{\varphi}

\newcommand{\ap}{\alpha}

\newcommand{\no}{\nonumber \\}

\newcommand{\ep}{\epsilon}

\begin{document}

\begin{titlepage}
\thispagestyle{empty}

\begin{flushright}
IPMU11-0071 \\
YITP-11-49
\end{flushright}

\begin{center}
\noindent{\Large \textbf{Probing AdS Wormholes by Entanglement Entropy}}\\
\vspace{2cm} \noindent{Mitsutoshi
Fujita\footnote[1]{e-mail: {\tt mitsutoshi.fujita@ipmu.jp}}, Yasuyuki
Hatsuda\footnote[2]{e-mail: {\tt hatsuda@yukawa.kyoto-u.ac.jp}} and Tadashi
Takayanagi\footnote[3]{e-mail: {\tt tadashi.takayanagi@ipmu.jp}}}\\
\vspace{1cm} {$^{1,3}$\it
Institute for the Physics and Mathematics of the Universe (IPMU), \\
 University of Tokyo, Kashiwa, Chiba 277-8582, Japan\\
} \vspace{5mm}

{$^2$\it
Yukawa Institute for Theoretical Physics, \\
Kyoto University, Kyoto 606-8502, Japan
}\\

\vskip 2em
\end{center}

\begin{abstract}
In this paper, we study the Lorentzian AdS wormhole solutions
constructed in arXiv:0808.2481 [hep-th]. 
Each of them is a classical solution interpolating between an AdS space and
a flat space in type IIB supergravity. 
We calculate the holographic entanglement entropy to
probe this geometry. Our analysis shows that there exits a mass gap
in its holographic dual gauge theory and that the entanglement
between the two boundaries is rather suppressed than that we naively expect
for wormholes. We also examine the holographic
conductivity on a probe D-brane in this spacetime.

\end{abstract}

\end{titlepage}

\newpage

\section{Introduction}

The AdS/CFT correspondence \cite{Maldacena} offers us a powerful way to study
quantum gravity on curved spacetimes in a non-perturbative way.
In particular, it will be intriguing to understand what the AdS/CFT
will predict about quantum gravity on topologically non-trivial spacetimes. In this
paper we would like to explore how we can apply the AdS/CFT to asymptotically AdS
spacetimes which look like Lorentzian wormholes, called AdS wormholes.

An AdS wormhole spacetime causally connects two different asymptotic
regions via a thin tube. Under a suitable energy condition in
Einstein gravity, it has been proved that there are no wormhole
solutions which connect two disconnected asymptotically AdS
boundaries \cite{GSWW}. However, recently wormhole-like solutions
have been found in \cite{Bergman:2008cv,Lu:2009ir,Wang:2009in} which
connect an asymptotically AdS boundary to an asymptotically flat
boundary.\footnote{There are other classes of AdS wormhole
solutions. We can find examples of the Euclidean AdS wormholes
\cite{WiYa,MaMa,AOP}. In Lorentzian spacetime, there exists AdS
wormholes in the Gauss-Bonnet gravity
\cite{Dotti:2006cp,Ali:2009ky,Arias:2010xg}.
Moreover, Lorentzian asymptotic AdS
spacetimes with multiple boundaries which are connected via horizons have been
studied in \cite{Skenderis:2009ju}. 
In this paper we will not discuss these.} These are smooth solutions in ten or eleven
dimensional supergravities. For example, a wormhole solution with a
D3-brane charge smoothly interpolates between AdS$_5\times$S$^5$ and
$\mathbb{R}^{1,9}$. This evades the no-go theorem because the asymptotically
AdS boundary appears only one side.

There is a well-known puzzle for an AdS wormhole \cite{WiYa,MaMa}. Since its boundary
consists of two disconnected manifolds, two dual CFTs live on them
according to the principle of AdS/CFT. Because the boundaries are
disconnected, we expect that the CFTs that live on them are
decoupled. However, in the AdS gravity picture, they are connected
by a throat and thus there should be non-zero interactions between
these two theories. In the black hole geometry, where the two
boundaries are connected by an event horizon and these two theories
are thermally entangled with each other \cite{MaE}. In the wormhole
case, the interpretation looks puzzling as the temperature of the
spacetime is vanishing.

Motivated by these, in this paper, we study the properties of AdS
wormhole solutions in order to understand their holographic duals.
In particular, we analyze the holographic entanglement entropy
\cite{RT,Hubeny:2007xt,NRT,CHM}. This is because the entanglement
entropy \cite{EE,Cardy,Review,Sol} is a rather general physical quantity
which can be defined in any quantum many body systems and is
useful to understand general properties of holography. Refer to
\cite{Hubeny:2007xt,Arias:2010xg} for earlier calculations of the entanglement entropy
of wormholes in different holographic setups.
Moreover, we will also analyze the holographic
conductivity as another probe of this spacetime.

This paper is organized as follows: In section two, we give a brief
review of the AdS wormhole solutions. In section three we calculate
the holographic entanglement entropy. In section four we compute the
holographic conductivity by using a probe D-brane. In section five,
we summarize conclusions.

\section{Brief Review of Lorentzian AdS Wormhole Solutions}

One of the most intriguing examples of the AdS wormhole solutions is
obtained in type IIB supergravity from a modification of
AdS$_5\times$S$^5$ solution \cite{Bergman:2008cv} \ba &&
ds^2=\left(\f{H(u)}{\cos
u}\right)^{1/2}\left(\f{a^2du^2}{16\cos^2u}+a^2d\Omega^2_5\right)
+H(u)^{-1/2}\left(\cos v(-dt^2+dz^2)+2\sin v dtdz +dx^2_1+dx^2_2
\right),\no && F_5=dt\we dz\we dx_1\we dx_2\we
d(H^{-1})+*\left(dt\we dz\we dx_1\we dx_2\we d(H^{-1})\right),
\label{AWsol}
 \ea
where we defined \be H(u)=\left(\f{l}{a}\right)^4\cdot
\left(\f{\pi}{2}-u\right), \ \ \ \ \
v=\s{\f{5}{2}}\left(\f{\pi}{2}-u\right),\ \ \ \ \
u=2\arcsin\( \frac{r(r^2+2a^2)^{1/2}}{\sqrt{2}(r^2+a^2)}\). \ee
Indeed, it is a simple
exercise to show that (\ref{AWsol}) satisfies the equation of motion
of type IIB supergravity in ten dimensions (here we set the string
coupling to be one $g_s=e^{\phi}=1$ just for a simple
normalization). The radial coordinate $u$ takes values between
$-\f{\pi}{2}<u<\f{\pi}{2}$. We can confirm that (\ref{AWsol}) is a
smooth solution which connects the AdS$_5\times$S$^5$ and the flat
space $\mathbb{R}^{1,9}$. In the limit $u\to\f{\pi}{2}$, the spacetime
approaches the AdS$_5\times$S$^5$ as \be ds^2_5\simeq
\f{l^2}{r^2}dr^2+\f{r^2}{l^2}(-dt^2+dz^2+dx_1^2+dx_2^2)+l^2d\Omega_5^2,
\ee where we used $u\simeq
\f{\pi}{2}-\f{a^4}{r^4}$ and $H\simeq \f{l^4}{r^4}$. On the other
hand, in the opposite limit $u\to -\f{\pi}{2}$, since $u\simeq
-\f{\pi}{2}+\f{a^4}{r^4}$ and $H\simeq
\f{l^4}{a^4}\pi-\f{l^4}{r^4}$, we obtain \be ds^2\simeq
\frac{\s{\pi}l^2}{a^2}(dr^2+r^2d\Omega^2_5)+\f{a^2}{\s{\pi}l^2}\left[\cos
v_0(-dt^2+dz^2)+2\sin v_0dtdz+dx_1^2+dx_2^2\right], \ee where we set
$v_0=\pi\s{5/2}$. It is obvious that this represents a flat
spacetime after an appropriate coordinate transformation. Through this analysis we also learned that the parameter
$l$ describes the AdS radius, while $a$ does the size of the
wormhole.

We can dimensionally reduce the this solution (\ref{AWsol}) to five
dimensions \cite{Bergman:2008cv}. The five dimensional gravity action
looks like \be S_5=\int
dx^5\s{-g}\left(R-\f{1}{2}\de_\mu\phi\de^\mu\phi
+\f{4}{l^2}(5e^{16\ap\vp/5}-2e^{8\ap\vp})\right), \ee where we set
$\ap=\s{5/48}$.
 \ba
&& ds^2_5=\left(\f{a}{l}\right)^{\f{10}{3}}\!\!\left[\f{a^2
du^2}{16\cos^2 u}\left(\f{H(u)}{\cos u}\right)^{\f{4}{3}}
\!+\!\f{H(u)^{\f{1}{3}}}{(\cos u)^{\f{5}{6}}}\left(\cos v
(-dt^2\!+\!dz^2)\!+\!2\sin v dtdz\!+\!dx_1^2\!+\!dx_2^2
\right)\right],\ \
\label{metric} \\
&& e^{-6\ap\vp/5}=\f{a^2}{l^2}\s{\f{H}{\cos u}}.
\ea

From the metric of this solution, we can show that there are no
time-like or null geodesic line which connects the two asymptotic
regions \cite{Lu:2008zs,Bergman:2008cv}. However, there is a non-geodesic 
time-like trajectory which connects them. This is the reason why the
authors of \cite{Bergman:2008cv} call this an AdS wormhole.

Another important property is the holographic stress energy tensor
for the background (\ref{AWsol}). The only non-trivial component
turns out to be \be T_{tz}=\f{\s{10}a^4}{8\pi l^5}. \ee Notice that
this breaks the standard energy conditions as the energy is
vanishing while the momentum is non-zero. However, this wormhole
solution is a smooth solution in type IIB supergravity and can be
embedded into string theory. Therefore it is still intriguing to
understand the structure of holography for this spacetime. This kind
of analysis will help us to understand holography in rather general
spacetimes, which is one of the most important problems in string
theory.

\section{Holographic Entanglement Entropy of AdS$_5$ Wormhole}

The entanglement entropy $S_A$ with respect to the subsystem $A$ is
defined by the von Neumann entropy $S_A=-\Tr \rho_A\log\rho_A$
for the reduced density matrix $\rho_A$. The reduced density matrix
is defined by tracing out the subsystem $B$, which is the complement
of $A$. In quantum field theories, we can specify the subsystem $A$
by dividing a time slice into two regions. In this paper we choose
the region for $A$ is given by a strip whose width is defined by
$L$. In order to understand properties of the AdS wormhole
solutions, we would like to study the behavior of the
entanglement entropy in this section.

In the holographic dual calculation \cite{RT,Hubeny:2007xt}, this quantity is 
given by the area of a codimension two extremal surface
$\gamma_A$ as \be S_A=\f{\mbox{Area}(\gamma_A)}{4G_N}.
\label{arealaw} \ee We require that the boundary of $\gamma_A$
coincides with that of the region $A$ and that $\gamma_A$ is
homotopic to $A$. If we have several extremal surfaces, then we pick
up the one with the smallest area. We will employ the five
dimensional metric (\ref{metric}) and analyze its three dimensional
minimal surfaces.

Below we consider two cases where the strip $A$ extends infinitely
(i) in the $x_1$ and $z$ and (ii) in the $x_1$ and $x_2$ directions.

\subsection{Case 1: Strip extended in the $x_1$ and $z$ directions}
Let us consider the holographic entanglement entropy for the
following subsystem $A$
\begin{align}
A=\{(x_1,x_2,z)|-\infty<x_1,z<\infty,\; -L/2<x_2<L/2\}.
\end{align}
By the translation symmetry, we can parameterize the surface as
\begin{align}
t=t(u),\qquad x_2=x_2(u).
\end{align}
The area functional is given by
\begin{align}
S=\( \frac{a}{l} \)^5 \int\! dx_1dz du \,\frac{H^{1/2}}{\cos^{5/4}u}\sqrt{L_1},
\end{align}
where
\begin{align}
L_1=\cos v\(\frac{a^2}{16}\frac{H}{\cos^{5/2}u}+(x_2')^2\)-(t')^2.
\end{align}
The equations of motion leads to the following two conserved quantities
\begin{align}
\frac{H^{1/2}\cos v x_2'}{\cos^{5/4}u \sqrt{L_1}}=\frac{1}{\alpha}, \quad
\frac{H^{1/2}t'}{\cos^{5/4}u \sqrt{L_1}}=\frac{1}{\alpha \beta},\label{eq:EOM1}
\end{align}
where $\alpha$ and $\beta$ are integration constants.
To define the entanglement entropy at a constant time (e.g.
$t=0$), we need to require that the subsystem $A$ is on the same
time slice: $T=2t(\pi/2)=0$.
Since we can show that solutions with $t' \ne 0$ and $T=0$ do not
exist, we focus on the solutions with $t'=0$ below. When $u$ takes
the minimal value $u_*$, $x_2'$ should diverge at this point. This
gives the relation between $u_*$ and $\alpha$,
\begin{align}
\alpha^2=f(u_*),\qquad f(u) \equiv \frac{\cos^{5/2}u}{H \cos v}.
\end{align}
One can easily solve \eqref{eq:EOM1},
\begin{align}
x_2(u)=\frac{a}{4}\int_{u_*}^u \frac{du}{\sqrt{(f(u_*)-f(u))\cos v}}.
\label{eq:sol1}
\end{align}
The length of the strip is given by $L=2x_2(\pi/2)$.
The area is given by
\begin{align}
{\rm Area}&=2\( \frac{a}{l} \)^5 \int\! dx_1dz du \,\frac{H^{1/2}}{\cos^{5/4}u}\sqrt{L_1} \nonumber \\
&=\frac{a}{2}\( \frac{a}{l} \)^5 L_{\perp}^2 \int_{u_*}^{\pi/2}du \frac{1}{f(u)}
\sqrt{ \frac{f(u_*)}{(f(u_*)-f(u))\cos v} }
\end{align}
where $L_\perp$ is the length in the $x_1$ and $z$ directions, and we assume $L_\perp \gg L$.
This area is still divergent because the integrand near $u=\pi/2$ behaves as
\begin{align}
{\rm Integrand} \sim \(\frac{l}{a}\)^4\frac{1}{(\frac{\pi}{2}-u)^{3/2}}.
\end{align}
To regularize the area,
we put a cut-off at $u=\pi/2-\epsilon$, then the regularized area behaves as
$a^2 L_\perp^2/(l\sqrt{\epsilon})+(\text{finite part})$. The divergent part is
in accord with the standard area law of the entanglement entropy \cite{EE}.
The cut-off independent finite area is given by
\begin{align}
A_{\rm fin}= \frac{a^2L_\perp^2}{l} \left[ \frac{1}{2}\int_{u_*}^{\pi/2}du \left\{\(\frac{a}{l}\)^4\frac{1}{f(u)}
\sqrt{ \frac{f(u_*)}{(f(u_*)-f(u))\cos v} }-\frac{1}{(\frac{\pi}{2}-u)^{3/2}}\right\}
-\frac{1}{\sqrt{\frac{\pi}{2}-u_*}}\right].
\end{align}

\begin{figure}[tb]
\begin{center}
\begin{tabular}{ccc}
\hspace{-3mm} \resizebox{50mm}{!}{\includegraphics{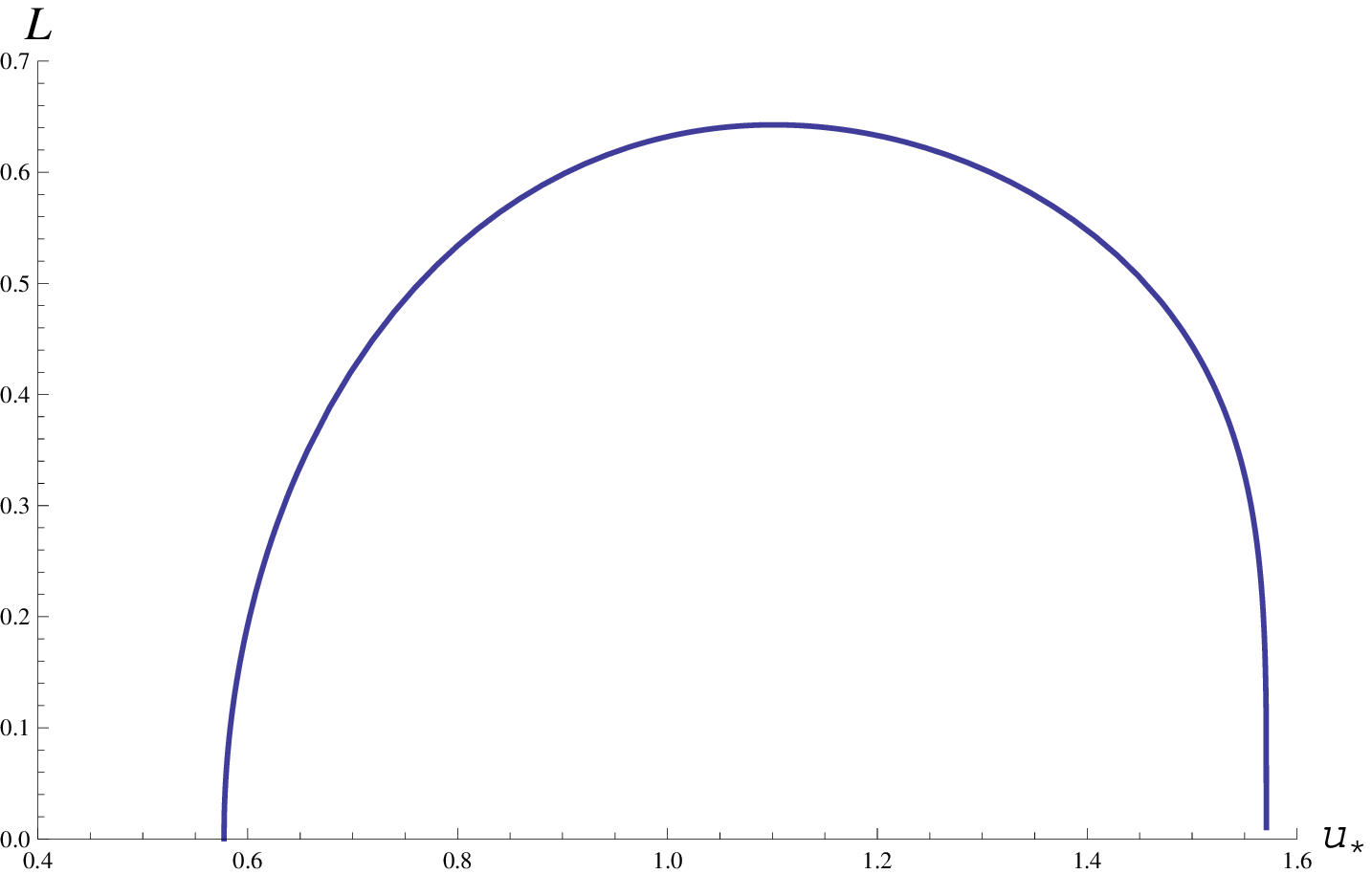}}
\hspace{-4mm} & \resizebox{60mm}{!}{\includegraphics{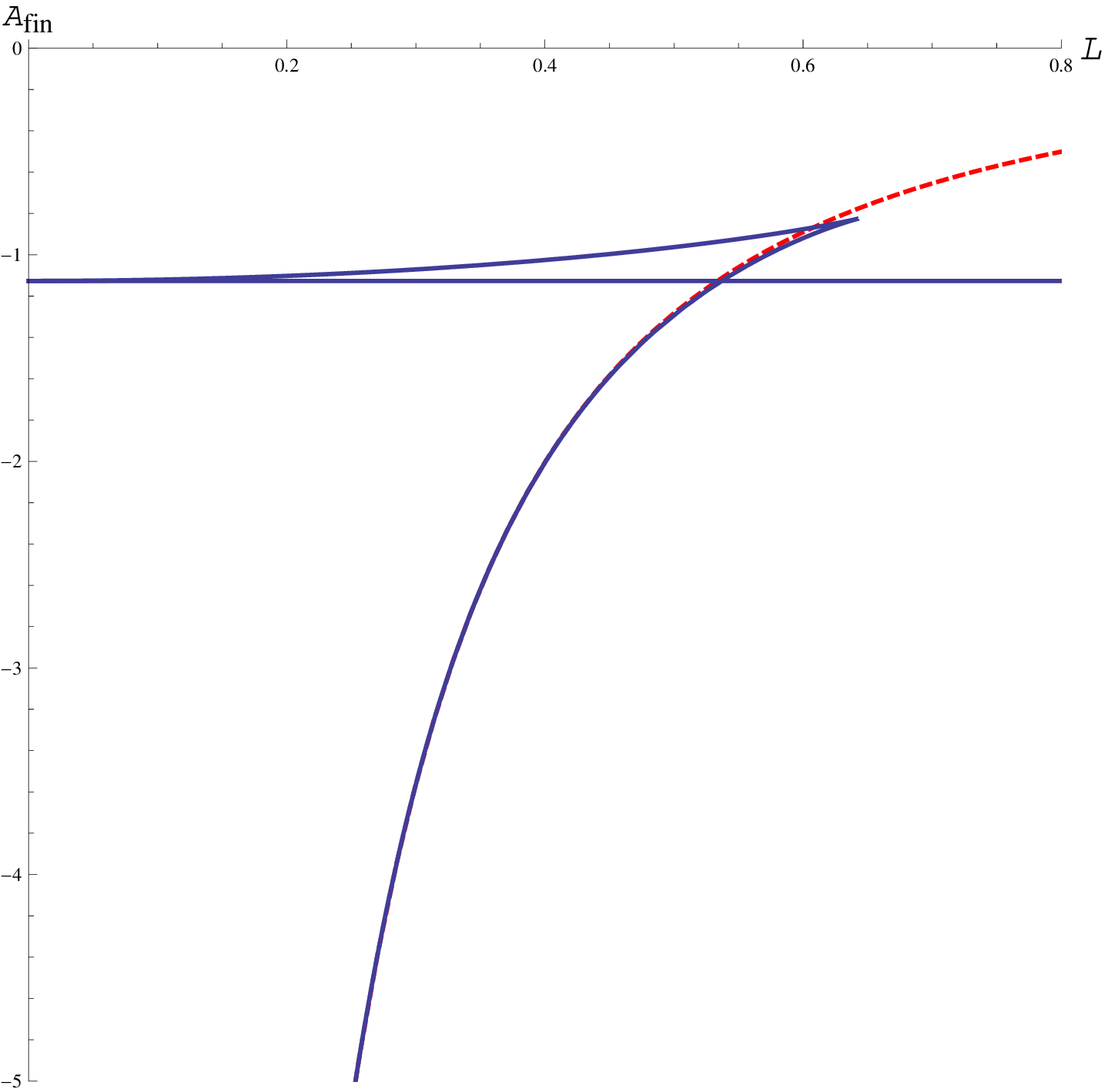}}
\hspace{-5mm} & \resizebox{50mm}{!}{\includegraphics{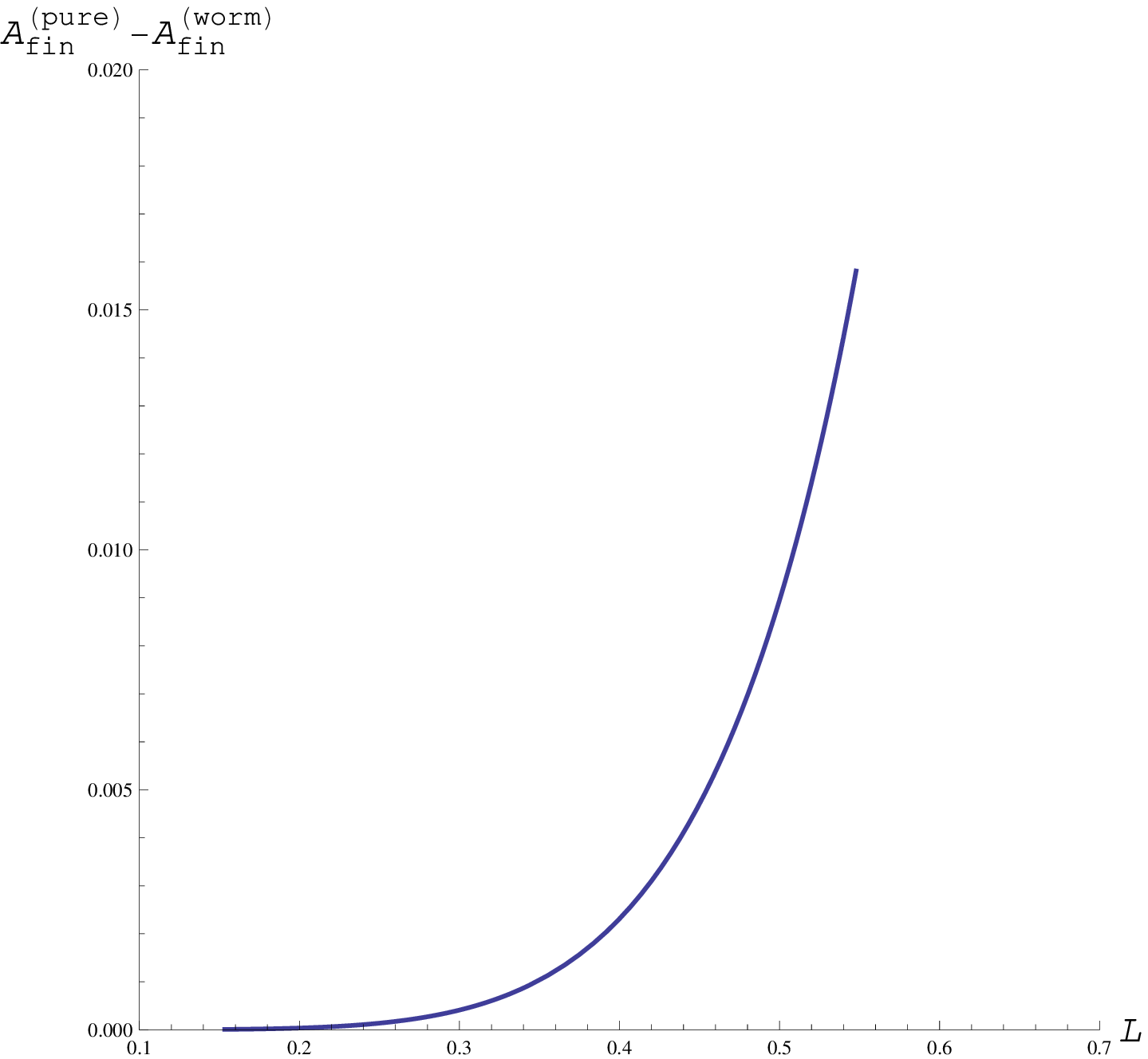}}
\\
(a) & (b) & (c)
\end{tabular}
\end{center}
\caption{(a) The $u_*$-dependence of $L$. $L$ becomes maximal at
$u_* \simeq 1.10$, and vanishes at $u_*=u_1, \pi/2$. We set $a=l=1$.
(b) The finite part of the area $A_{\rm fin}$ as a function of $L$. The
red dashed graph is the one for the pure AdS. The blue solid one
corresponds to the AdS wormhole. We set $a=l=L_\perp=1$. (c) The difference between the
holographic entanglement entropy (i.e. $A^{\rm (pure)}_{\rm fin}-A^{\rm (worm)}_{\rm fin}$) 
for the pure AdS and that for the AdS
wormhole.} \label{fig:graph1}
\end{figure}

The $u_*$-dependence of length $L$ is shown in Fig.~\ref{fig:graph1}(a).
This figure shows that $L$ has a bound at some value of $u_*$,
and the solution exists only for $u_1\leq u_* \leq \pi/2$ where $u_1$ is the value such that $\cos v(u_1)=0$, that is,
\begin{align}
u_1=\frac{\pi}{2}\(1-\frac{\sqrt{10}}{5}\) \simeq 0.577338.
\label{eq:u1}
\end{align}
Therefore for a small enough value of $L$ we have two connected
minimal surfaces.

Moreover, for any value of $L$, there are disconnected surfaces
given by $x_2=L/2$ and $x_2=-L/2$, extending from $u=\pi/2$ to
$u=u_1$. At $u=u_1$, the surface can end as the area density gets
vanishing as follows from the metric (\ref{metric}).

In this way, we have three branches as a function of $L$: two are
connected and one is disconnected. In Fig.~\ref{fig:graph1}(b), we
plot the relation between $L$ and the finite part of
the area for each surface. Among the three surfaces, we need to
select the one with the smallest area. First we can ignore the
branch which always takes the largest value. The remaining two
branches actually compete with each other: one takes a constant
value and the other one has a concave shape. It is important that
the constant one intersects with the concave one at some value of
$L$ (see Fig.~\ref{fig:graph1}(b)). This means that for $L$ larger
than this value $L_c$, the surface that consists of two disconnected
planes $x_2=\pm L/2$ ($u_1\leq u \leq \pi/2$) and the plane $u=u_1$
($-L/2\leq x_2 \leq L/2$) has the smaller area than the surface of
the solution \eqref{eq:sol1}. On the other hand, when $L<L_c$ we
need to choose the concave one. Thus we find a sort of phase
transition at $L=L_c$.

Qualitatively, this behavior looks very similar to the one obtained
for gravity duals of confining gauge theories
\cite{NiTa,KKM,PaPa}, where an analogous phase transition occurs and
corresponds to the confinement/deconfinement transition 
(a similar behavior has recently been
confirmed by calculations in lattice gauge theories \cite{Lata,Latb,Latc}).
This implies that the dual CFT has a mass gap, which will probably
correspond to some confining gauge theory. Indeed, we can confirm
that the holographic entanglement entropy for the AdS$_5$ wormhole
is smaller than that for the pure AdS$_5$ as plotted in
Fig.~\ref{fig:graph1}(c).

\subsection{Case 2: Strip extended in the $x_1$ and $x_2$ direction}
Let us next consider the subsystem
\begin{align}
A=\{(x_1,x_2,z)|-\infty<x_1,x_2<\infty,\; -L/2<z<L/2\}
\end{align}
If we parameterize the surface as
\begin{align}
t=t(u),\qquad z=z(u),
\end{align}
the action is given by
\begin{align}
S=\( \frac{a}{l} \)^5 \int\! dx_1dx_2 du \frac{H^{1/2}}{\cos^{5/4} u} \sqrt{L_2},
\end{align}
where
\begin{align}
L_2 = \frac{a^2}{16}\cdot \frac{H}{\cos^{5/2} u}+\bigl(-(t')^2+(z')^2\bigr)\cos v+2 t' z'\sin v.
\end{align}
The equations of motion lead to 
\begin{align}
\frac{H^{1/2}}{\cos^{5/4}u}\frac{z' \cos v+t' \sin v}{\sqrt{L_2}}=\frac{1}{\alpha},\quad
\frac{H^{1/2}}{\cos^{5/4}u}\frac{z' \sin v-t' \cos v}{\sqrt{L_2}}=\frac{1}{\beta},
\end{align}
where $\ap$ and $\beta$ are constants.
The solutions are given by
\begin{align}
t(u)=\frac{a}{4}\int_{u_*}^u du \frac{C \sin v-\cos v}{\sqrt{g(u_*)-g(u)}},\quad
z(u)=\frac{a}{4}\int_{u_*}^u du \frac{C \cos v+\sin v}{\sqrt{g(u_*)-g(u)}},
\label{eq:sol2}
\end{align}
where $C \equiv \beta/\alpha$ and
\begin{align}
g(u)\equiv\frac{2C \sin v-(1-C^2)\cos v}{H}\cos^{5/2}u.
\end{align}
The minimal value $u_*$ is related to $\alpha$ and $\beta$ as
\begin{align}
\alpha^2\beta^2 H_*+[ (\alpha^2-\beta^2)\cos v_*-2\alpha \beta \sin v_*]\cos^{5/2}u_*=0
\end{align}

As in the previous subsection, we impose the constant time slice condition $T=2t(\pi/2)=0$.
This constraint relates the constant
$C$ to $u_*$. The area is given by
\begin{align}
{\rm Area}&=2\( \frac{a}{l} \)^5 \int\! dx_1dx_2 du \frac{H^{1/2}}{\cos^{5/4} u} \sqrt{L_2}\nonumber \\
&=\frac{a}{2}\( \frac{a}{l} \)^5 L_\perp^2 \int_{u_*}^{\pi/2}du \frac{H}{\cos^{5/2}u}
\sqrt{ \frac{g(u_*)}{g(u_*)-g(u)} }
\end{align}
As well as before, we define the finite part of the area by
\begin{align}
A_{\rm fin}\equiv\frac{a^2L_\perp^2}{l} \left[ \frac{1}{2}\int_{u_*}^{\pi/2}du\left\{ \(\frac{a}{l}\)^4\frac{H}{\cos^{5/2}u}
\sqrt{ \frac{g(u_*)}{g(u_*)-g(u)}} -\frac{1}{(\frac{\pi}{2}-u)^{3/2}}\right\}
-\frac{1}{\sqrt{\frac{\pi}{2}-u_*}}\right].
\end{align}

\begin{figure}[tb]
\begin{center}
\begin{tabular}{cc}
\hspace{-3mm}
\resizebox{70mm}{!}{\includegraphics{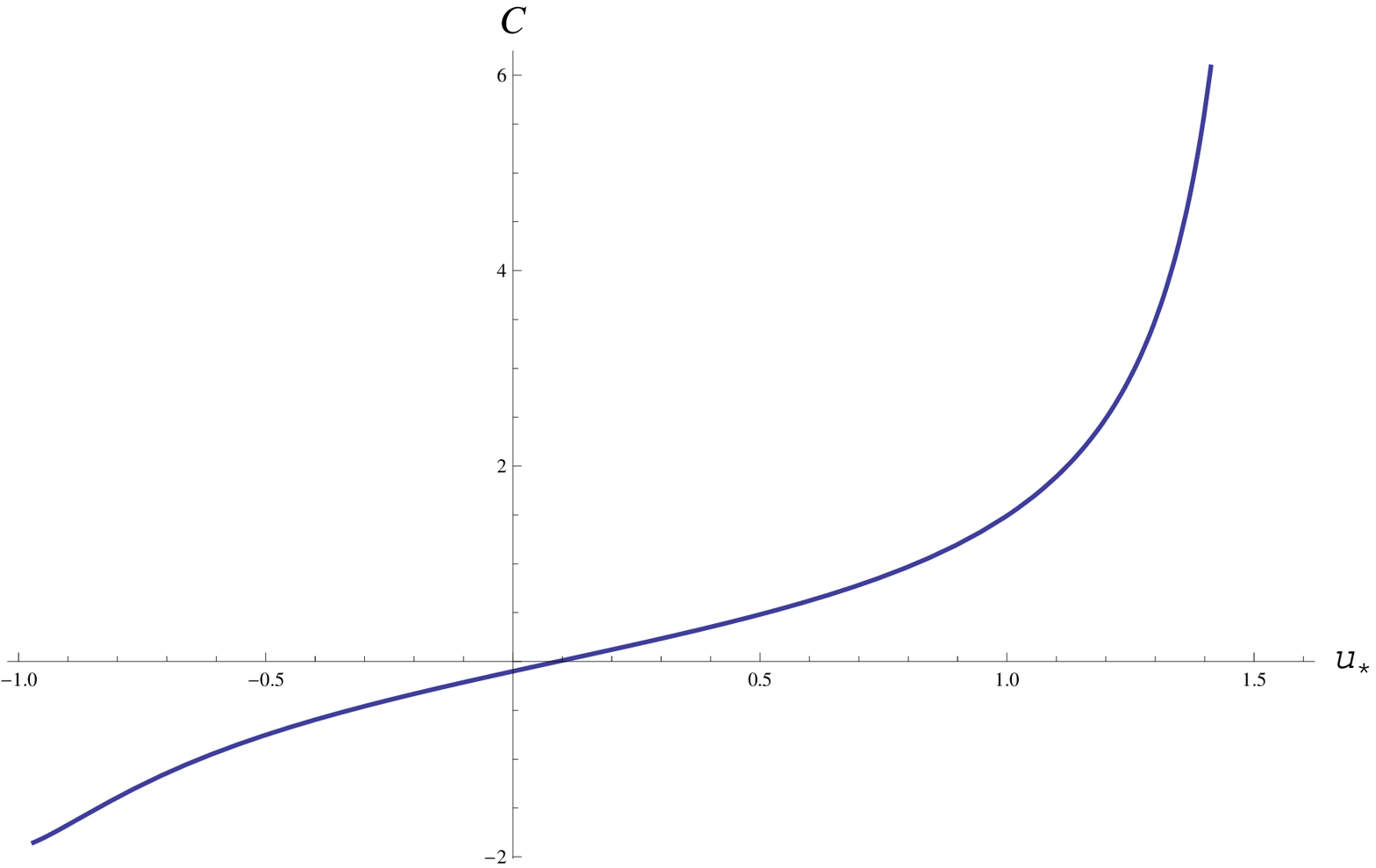}}
\hspace{-4mm}
&
\resizebox{70mm}{!}{\includegraphics{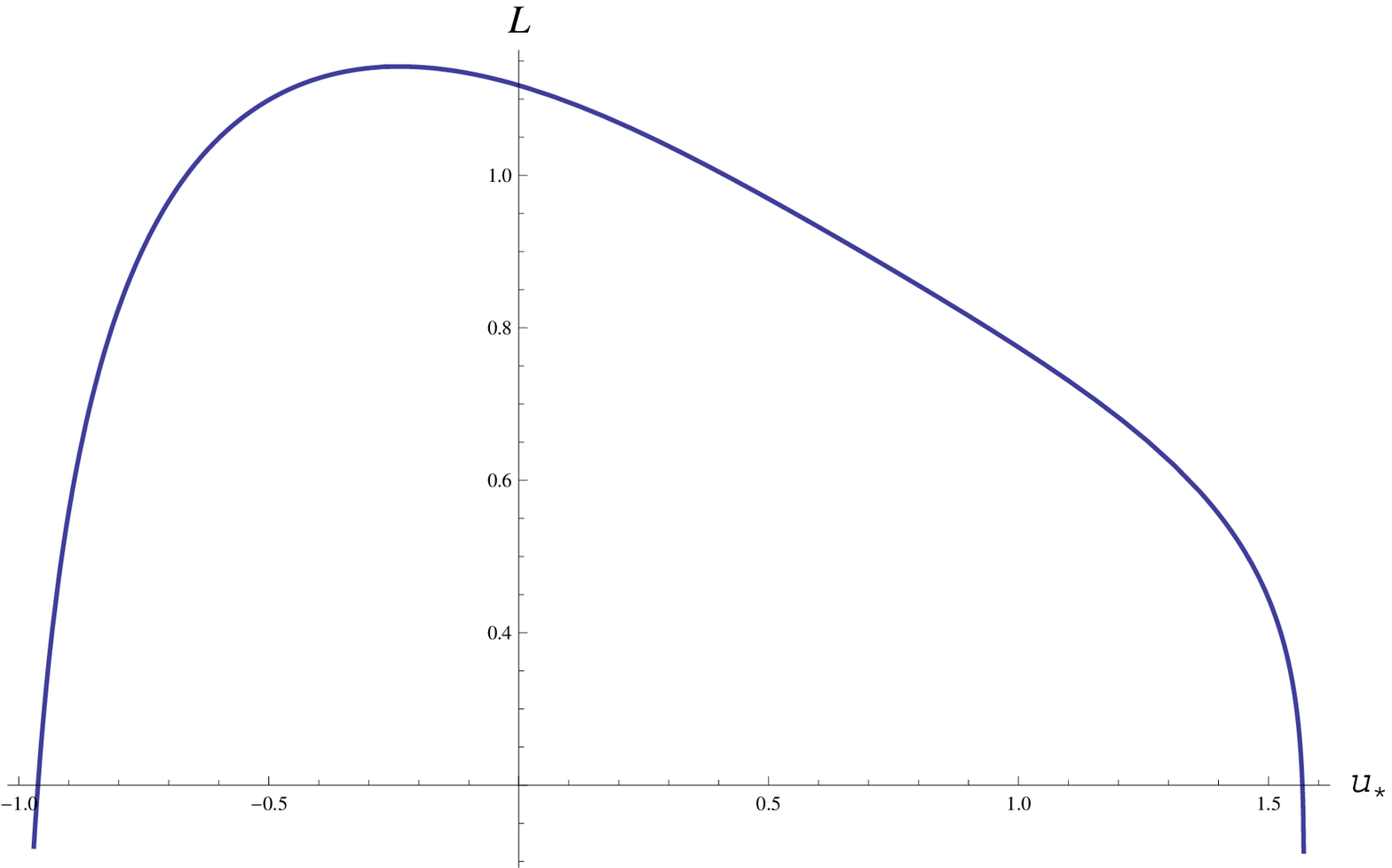}}
\hspace{-5mm}
\\
(a) & (b)
\end{tabular}
\end{center}
\caption{The $u_*$-dependence of (a) $C$ and (b) $L$.}
\label{fig:graph2}
\end{figure}

\begin{figure}[tb]
  \begin{center}
\begin{tabular}{ccc}
\hspace{-3mm} \resizebox{50mm}{!}{\includegraphics{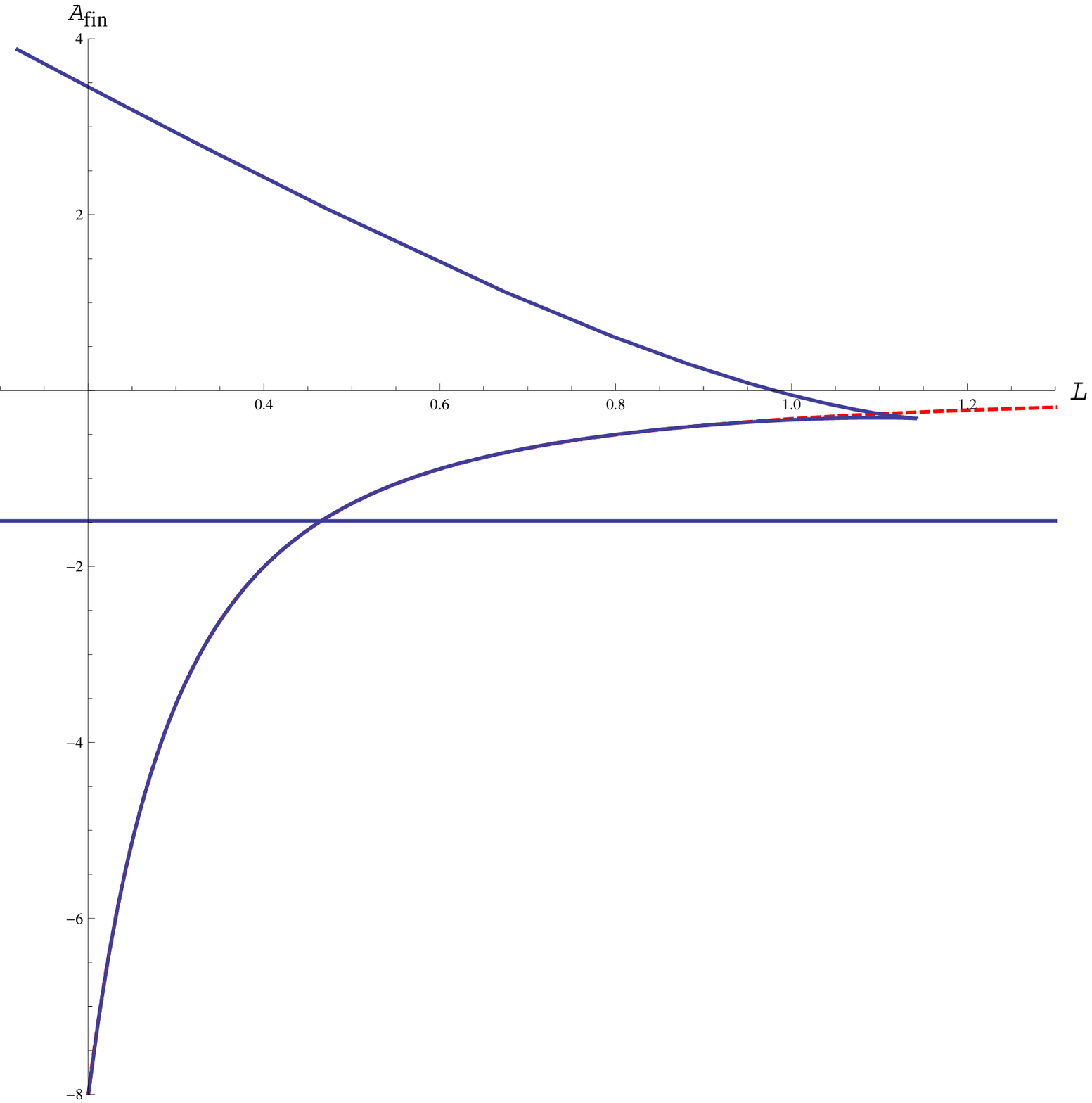}}
\hspace{-3mm} & \resizebox{50mm}{!}{\includegraphics{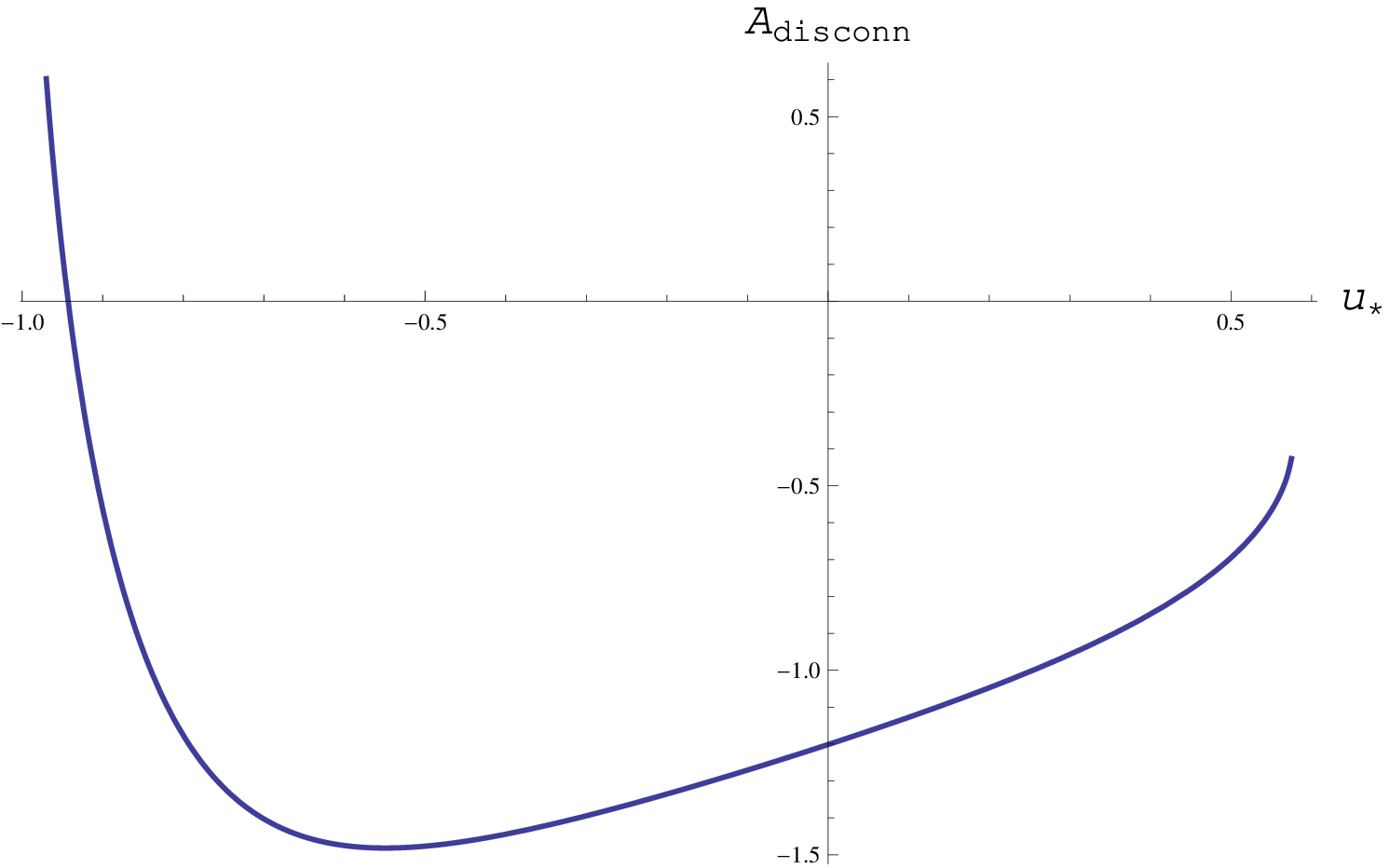}}
\hspace{-4mm} & \resizebox{45mm}{!}{\includegraphics{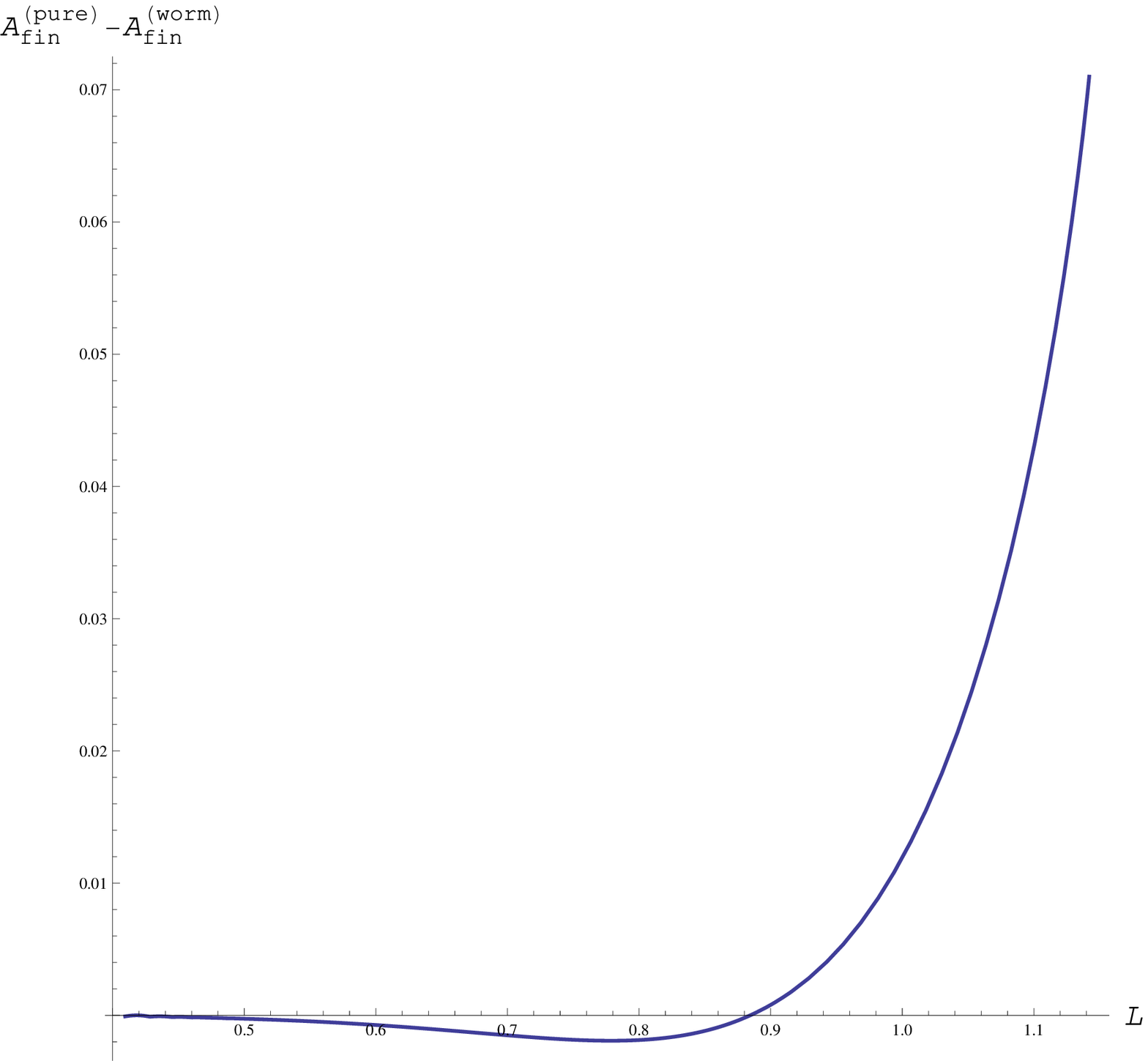}}
\hspace{-5mm}
\\
(a) & (b) & (c)
\end{tabular}
  \end{center}
  \caption{(a) The plot of $L$ versus the finite area. (b) The $u_*$-dependence of the area of the disconnected
  surface extending to the range $u_1\leq u \leq \pi/2$. (c) The difference between the
  holographic entanglement entropy for the pure AdS and that for AdS wormhole. }
  \label{fig:graph3}
\end{figure}

Once we find $C=C(u_*)$ from the condition $T=0$,  we can evaluate the length
$L=2z(\pi/2)$ and the area as functions of $u_*$. We plot the
$u_*$-dependence of $C$ and $L$ in Fig.~\ref{fig:graph2}(a) and (b),
respectively, and the relation between $L$ and the finite part of
the area is shown in Fig.~\ref{fig:graph3} (a). The area again has
three branches.

As in the previous subsection, the above connected surfaces are
defined only for small enough values of $L$ and we can find a
disconnected surface defined for any $L$.
This disconnected surface extends to the range $u_1\leq u \leq \pi/2$ since
the area density at $u=u_1$ is vanishing as mentioned before.
Among the connected surfaces constructed in \eqref{eq:sol2} with the constraint $T=0$,%
\footnote{The surfaces with $T\ne 0$ have the larger areas than those with $T=0$,
or do not satisfy the null expansion conditions required in \cite{Hubeny:2007xt}.}
we compute the area of the part $u_1 \leq u \leq \pi/2$ as a function of $u_*$, and
find that the area takes a minimal value $\simeq -1.48 a^2L_\perp^2/l$ at $u_*\simeq -0.549$
(see Fig.~\ref{fig:graph3}(b)).
Therefore the disconnected surface that has the minimal area is given by the part
$u_1 \leq u \leq \pi/2$ of the connected one for $u_*\simeq -0.549$.

In summary, for this choice of the subsystem, we again encounter a
behavior similar to confining gauge theories \cite{NiTa,KKM,PaPa}. For large
enough $L$, our entanglement entropy is smaller than that for the
pure AdS as shown in Fig.\ref{fig:graph3} (c) and this again
suggests that the theory has a mass gap.

One may naturally think that the wormhole geometry suggests that the
two holographic duals on the two boundaries are entangled with each
other. If this is true, then the entanglement entropy should be
greater than that for the pure AdS space. Indeed, as we explain in
the appendix A, we can obtain this behavior for an geodesically 
traversable AdS wormhole (\ref{wma}). The amount of
the increased entropy is extensive and therefore the holographic dual
looks like a thermal state. However, the metric (\ref{wma}) cannot
be embedded into Einstein gravity without violating the null energy
condition of matter fields \cite{GSWW}.

On the contrary, our holographic results for the AdS wormhole
solution (\ref{metric}), which can be realized in type IIB
supergravity \cite{Bergman:2008cv}, show that the entropy is reduced
as opposed to this naive guess. Also our analysis reveals that the
entanglement entropy between the two boundary theories is vanishing.
All these tell us that the presence of two boundaries does not cause
the quantum entanglement between the two dual theories at least from
the viewpoint of the entanglement entropy. It may be possible that
this slightly surprising property is generally true in the
`wormhole' gravity duals in physically sensible gravity theories.
Finally we note that the AdS$_3$ wormhole solutions constructed in \cite{Lu:2008zs, Bergman:2008cv}
also have the same property.

\section{Holographic Conductivity}

To probe the AdS wormhole, next we would like to study
the holographic conductivity.
 We calculate the AC conductivity on
 a probe D3-brane in the AdS$_5$ wormhole solution (\ref{AWsol}).
For simplicity, we set $a=l=1$. The worldvolume coordinates of the
probe D3-brane are given by $(t,z,u,x_1)$. The brane is localized in
the $x_2$ and $S^5$ direction. The induced metric of the D3-brane
then becomes asymptotically $AdS_4$ in the limit $u\to\pi/2$.

Another motivation of this setup is a possibility of constructing
a holographic dual of a system with impurities (for earlier works
from different viewpoints see \cite{Hartnoll:2008hs,Fujita:2008rs,KKY,JH,AY,RTU}).
 We regard the theory that lives on the asymptotically flat
 boundary as the one for the impurities sector, while the main quantum system
 lives on the AdS boundary.  Assuming that these
 impurities only absorb the energy from the outside, we can impose
in-going boundary conditions as we will see below. However, notice that we can also impose
the Dirichlet boundary condition for which the system looks like a insulator, though we will not
discuss this in detail.

\subsection{Vanishing Charge Density}

We first consider the case where the charge density on the brane is
vanishing.  In the quadratic order of the worldvolume $U(1)$ fields,
the Lagrangian is given by the Maxwell theory \be L_{D3}=-\int
d^4x\dfrac{1}{4}\sqrt{-g}F^{\alpha\beta}F_{\alpha\beta}, \ee where
$\sqrt{-g}=1/(4H^{1/2}\cos^{5/4}(u))$ and
 $\alpha,\beta =t,x,z,u$ $(x=x_1)$. Here, the gauge field is normalized properly.

Imposing the gauge fixing $A_{u}=0$, the equations of motion for the
vector field $A_{x}=a_{x}(u)e^{-i\omega t}$ become the
following differential equation\footnote{In this paper, we do not analyze other vector fields $A_t$ and $A_z$ since for $A_x$, the analysis with finite charge is simpler than that for $A_t$ and $A_z$. } :
\ba
-\omega^2 \sqrt{-g}g^{tt}g^{xx}A_{x}+\partial_u (\sqrt{-g}g^{uu}g^{xx}\partial_uA_{x})=0. \label{EOM14}
\ea

In the asymptotically $AdS_4$ space, the holographic AC conductivity
is given as follows. The metric near the boundary $w\to 0$ is
given by \be ds^2=\dfrac{dw^2-dt^2+d\vec{x}^2}{w^2}+..... \ee Then,
the AC conductivity is described using the asymptotic values of
$A_x$ as follows:
\begin{align}
\sigma_{xx}=\dfrac{j_x}{E_x}=\dfrac{A_x^{(1)}}{-i\omega
A_x^{(0)}},\quad
 A_x(w)=A_x^{(0)}+wA_x^{(1)}+... \label{CON438}
\end{align}

Substituting the $AdS_5$ wormhole metric into \eqref{EOM14}, the following differential equations are obtained:
\ba
\partial_u^2 A_{x}+\partial_u\log\Big(\dfrac{4\cos^{5/4}(u)}{H^{1/2}}\Big)\partial_u A_{x}+\omega^2\dfrac{\cos (v) H}{16\cos^{5/2}(u)}A_{x}&=0. \label{EOM439}
\ea

\begin{figure}[tb]
   \begin{center}
\begin{tabular}{cc}
\hspace{-3mm} \resizebox{60mm}{!}{\includegraphics{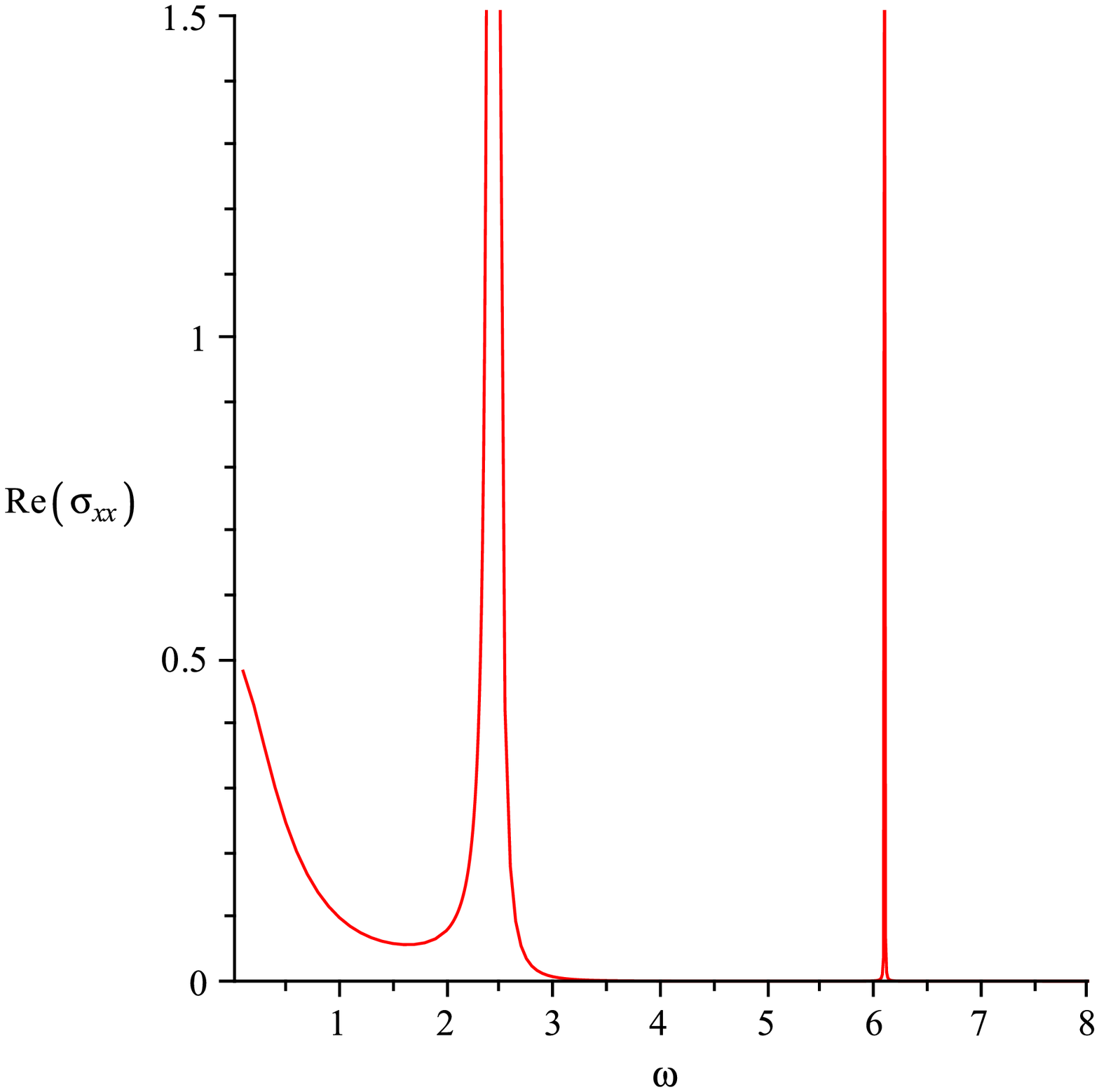}}
\hspace{-4mm} & \resizebox{60mm}{!}{\includegraphics{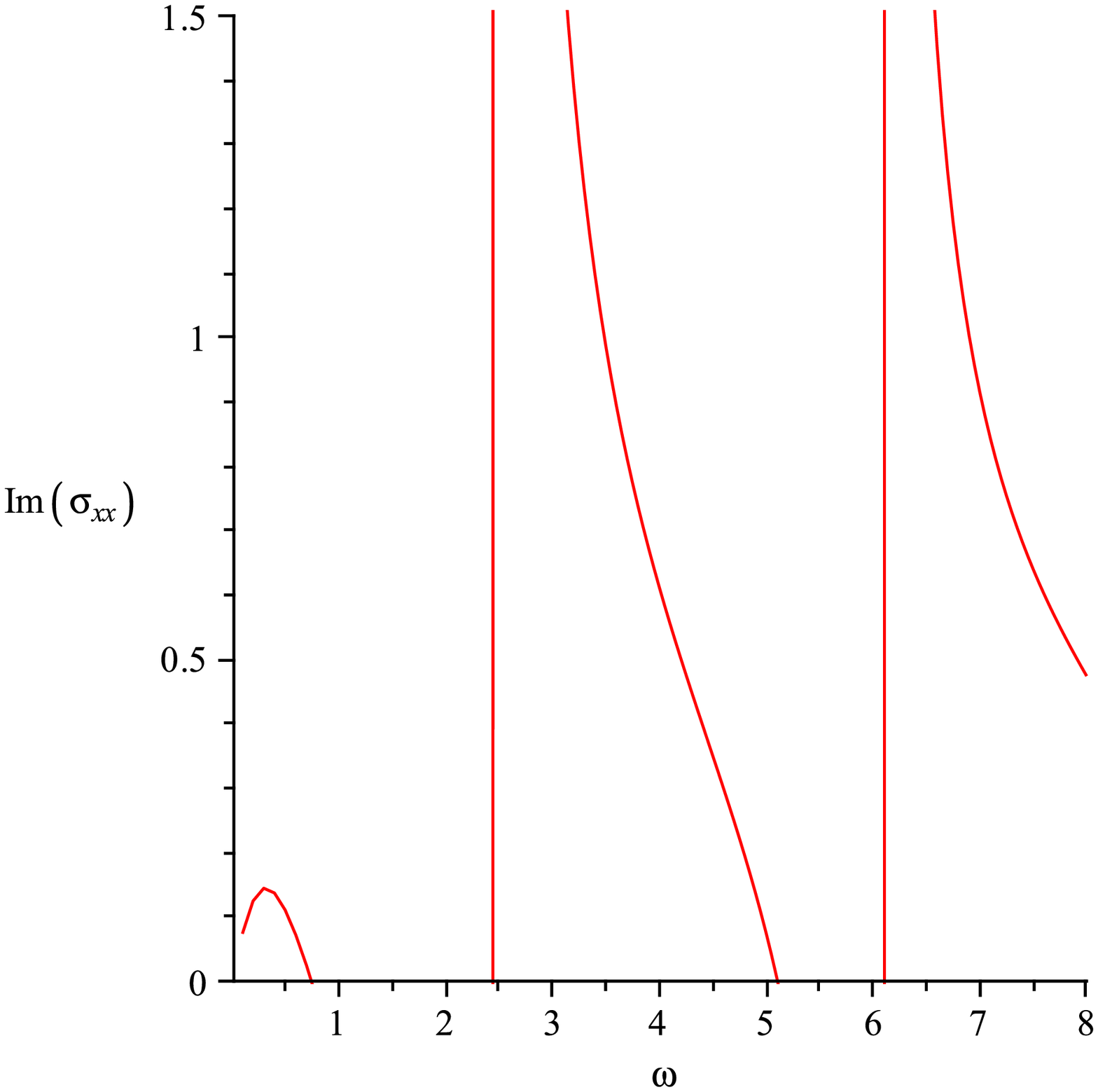}}
\\
(a) & (b)
\end{tabular}
   \end{center}
   \caption[.]{The $\omega$-dependence of (a) $\re \sigma_{xx}$ and (b) $\im \sigma_{xx}$.}
\label{cx}
\end{figure}

The solutions for the differential equation \eqref{EOM439}
behave as follows at the $AdS$ boundary $u\to \pi/2$ (we defined
$\eta =\pi/2 -u$) \be A_{x}\sim e^{\pm i\omega
\eta^{1/4}}=e^{\pm i\omega w},\quad w\sim \eta^{1/4}, \label{BOU112}
\ee where $w$ is introduced to use the formula \eqref{CON438}. Then, the derivative in terms of $w$ can be
rewritten as
 \be
 \partial_w A_{x}|_{w\to 0}=A_{x}^{(1)}=-4\eta^{3/4}\partial_uA_{x}|_{u=\pi/2}.
 \ee
 We require that $A_{x}$ satisfies the incoming boundary condition at the flat boundary for
 $u =-\pi/2$.  Rewriting
 the radial coordinate $u=-\pi/2 +\delta$ and imposing the ingoing boundary condition, the solution
  in the limit $u =-\pi/2$ should behave as
 \be
 A_{x}\sim e^{\pm i\sqrt{\pi \cos v_0}\omega \delta^{-1/4}}, \label{BOU442}
 \ee
 where $v_0$ is the value of $v$ at $u=-\pi/2$.
  To compute the holographic AC conductivity, we impose the ingoing 
  boundary condition \cite{Son:2002sd,
Policastro:2002se} for $A_{x}$ at $u\to -\pi/2$, namely the plus sign for \eqref{BOU442}.
Then compute the AC conductivity using the formula in
\eqref{CON438}. We plot $\omega$-dependence of the real part and
the imaginary part of the conductivity $\sigma_{xx}$  in Fig.
\ref{cx} (a) and (b).

 Notice that in both cases, the DC conductivity is non-vanishing
 as in the probe D-brane in AdS black holes \cite{KB}.
 In particular, we can find a Drude peak for $\sigma_{xx}$. It is
 important that we get this behavior in spite of the absence of a
black hole horizon in our wormhole spacetime. This suggests our
holographic calculation by imposing the in-going boundary condition
at the second boundary captures thermal behavior of the metallic
system. The several peaks of the real part of the conductivity
correspond to the resonances dual to mesonic bound states in the
confining geometry as similar to the result in \cite{NiRT}.

\subsection{Finite density}

Finally we compute the conductivity $\sigma_{xx}$ at
finite density. Consider the background of $A_t$ which gives
the electric charge density in the dual theory.

The DBI action with the worldvolume $U(1)$ gauge field is given by
\begin{align}
L_{D3}&=-T_{D3}\int d^4x\sqrt{-\det (g+F)} \nonumber \\
&=-T_{D3}\int
d^4x\sqrt{-g}\sqrt{1+g^{uu}g^{tt}F_{ut}^2+g^{tt}g^{xx}F_{tx}^2+g^{uu}g^{xx}F_{ux}^2},
\label{ACT115}
\end{align}
where we defined 
$T_{D3}=1/((2\pi)^3g_sl_s^4)$ and $g=\det (g_{\mu\nu})$. We simply set $T_{D3}=1$ below.

To obtain the conductivity at the finite density, we first determine
the charged background of $A_t$ and then we analyze the fluctuation
of $A_x$ to the quadratic order of $A_x$. The solution $A_t'$ is
given by \be
A_t'=\dfrac{\rho}{\sqrt{-(g(g^{tt}g^{uu})^2+g^{tt}g^{uu}\rho^2)}},
\label{EOM118} \ee where $\rho$ is a constant proportional to the
charge density and $A_t'=\partial_u A_t$.

Substituting the solution \eqref{EOM118} with $\rho\neq 0$  into
\eqref{ACT115} and expanding the action in terms of $A_x$, the
following quadratic action is obtained: \be -\dfrac{1}{2}\int d^4x
\sqrt{-g\Big(1+\dfrac{\rho^2}
{gg^{uu}g^{tt}}\Big)}(g^{tt}g^{xx}F_{tx}^2+g^{uu}g^{xx}F_{ux}^2).
\label{ACT119} \ee In the $A_u=0$ gauge, the equation of motion for
$A_x=A_x(u)e^{-i\omega t}$ is derived from \eqref{ACT119} as
follows: \be A_x''+\log
'\Big(4\cos^{5/4}(u)\sqrt{\dfrac{1}{H}+\dfrac{\rho^2}{\cos
v}}\Big)A_x'+\omega^2\dfrac{\cos(v)H}{16\cos^{5/2}(u)}A_x=0.
\label{EOM113} \ee Note that there is a singularity at $\cos(v)=0$
(i.e. $u=u_{1}=\pi/2-\pi/\sqrt{10}$) in \eqref{EOM113}, as opposed
to the previous case of $\rho=0$. This singularity appears because
of the F-string charge induced on the probe D3-brane with the
electric flux $F_{ut}=A'_t$. Indeed, if we consider an F-string which
simply extends in $u$ direction, the induced metric of this F-string
becomes singular at $u=u_1$. See also the D3/D7 brane system with the electric flux~\cite{Kobayashi:2006sb}.

\begin{figure}[tb]
   \begin{center}
\begin{tabular}{cc}
\hspace{-3mm} \resizebox{60mm}{!}{\includegraphics{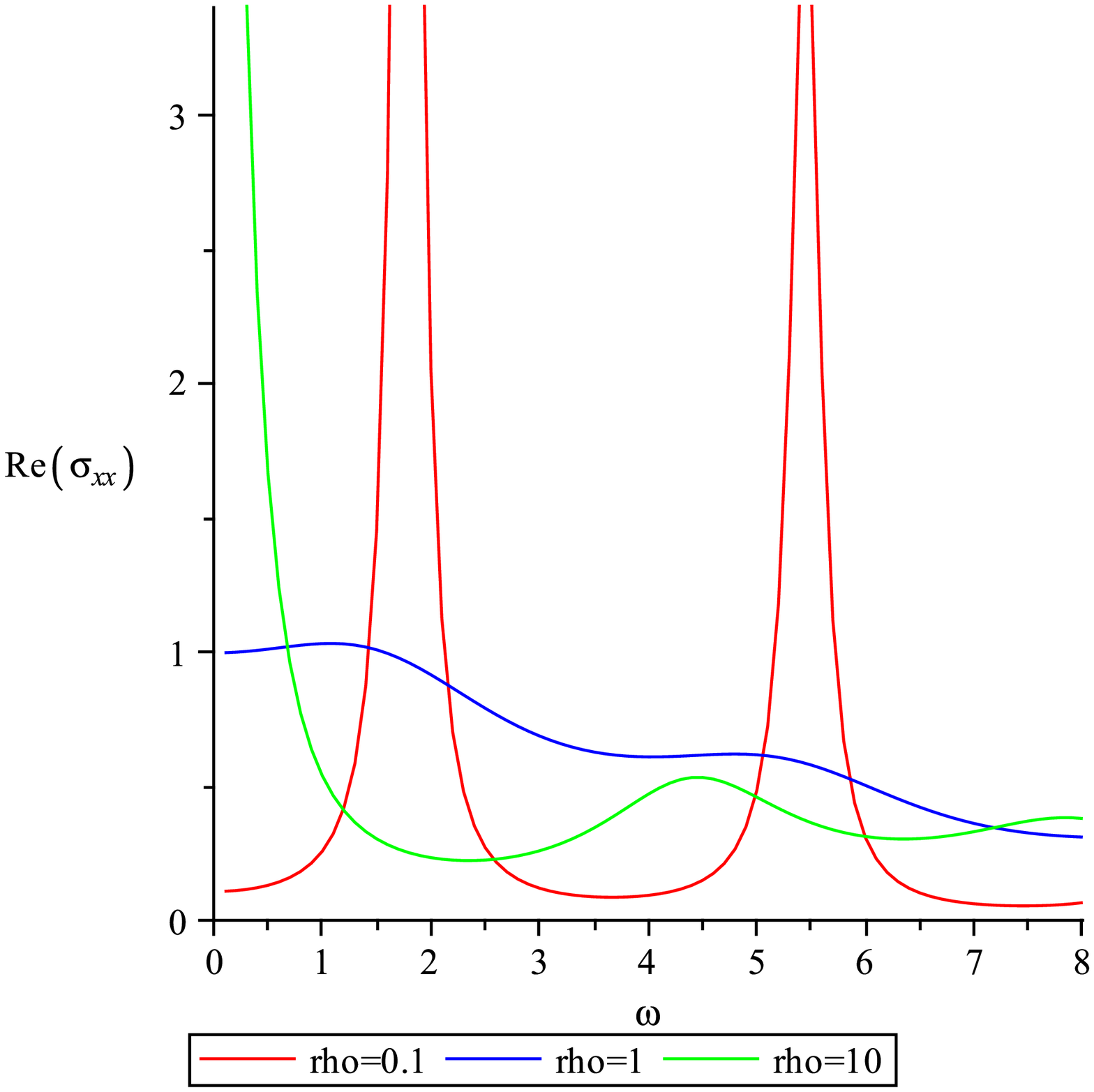}}
\hspace{-4mm} &
\resizebox{60mm}{!}{\includegraphics{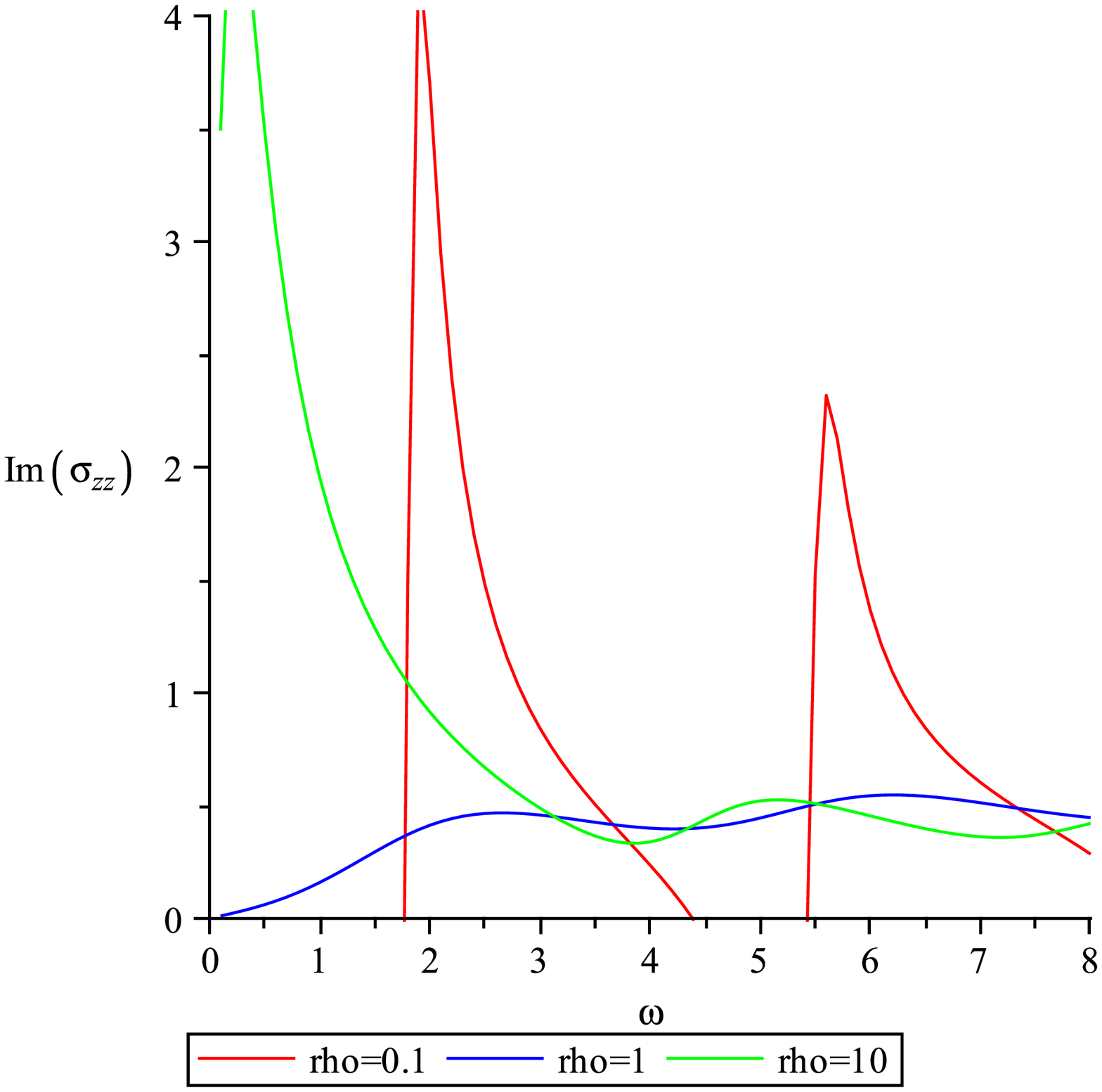}}
\\
(a) & (b)
\end{tabular}
   \end{center}
   \caption[.]{The $\omega$-dependence of (a) $\re \sigma_{xx}$ and
   (b) $\im \sigma_{xx}$
   at finite density $\rho=0.1,1$ and $10$ is plotted. }
\label{Redens}
\end{figure}

Therefore, a boundary condition should be imposed at $u=u_1$. The
solution of \eqref{EOM113} near $u=u_1$ behaves as \be A_x\sim \exp
\left(\pm i {\frac {  (u-u_1)^{3/2}\sqrt {2\pi}\omega }{ 12 \sin
^{5/4} \left( \pi/\sqrt{10} \right)}} \right). \label{AXX122} \ee To
impose the incoming boundary condition, we should choose the minus
sign in \eqref{AXX122}. Since the behavior of the solution at the
boundary $u=\pi/2$  behaves as \eqref{CON438} and \eqref{BOU112}, we
can compute the AC conductivity similar to the
$\rho=0$ case. We plot $\omega$-dependence of the real part and the
imaginary part of the conductivity $\sigma_{xx}$ at the finite
density in Fig. \ref{Redens} (a) and (b). Indeed, this has the
expected feature that the DC conductivity increases as the charge
density $\rho$ does. Note that in Fig. \ref{Redens} (a),  the Drude peak almost vanishes for the case $\rho = 0.1$ which is not similar to the case $\rho=0$ (Fig. \ref{cx}). This difference appears since the boundary condition is different between them. Though we do not intend to conclude which boundary condition is correct in this paper, it is natural to choose the boundary condition at $u=u_1$ for the continuity about $\rho$.

\section{Conclusions}

In this paper, we studied the properties of AdS wormhole solutions
which connect the asymptotically AdS boundary to the flat one. We
concentrated on the example of AdS$_5$ wormhole constructed in
\cite{Bergman:2008cv}. Even though its holographic stress tensor
suggests an unusual gauge theory dual, 
it is a smooth solution in type IIB
supergravity and it is instructive to understand its holography.

We calculated the holographic entanglement entropy in order to probe
the structure of this type of the spacetime as well as to better
understand its holographic dual. We may naively think that the
presence of the two boundaries suggest the quantum entanglement
between two holographic dual theories live on each of the
boundaries. However, what we actually find is not like this. The
entanglement entropy is generically reduced by the presence of the
second asymptotic region and shows the behavior common to mass
gapped theories. We observed a phase transition analogous to the
confinement/deconfinement transition when we change the size of the
subsystem. This means that the two asymptotic regions are separated
from each other from the viewpoint of the entanglement entropy. This
may partially resolve the original problem of the AdS/CFT for
wormholes because the entanglement between the two theories, which
looks confusing on the CFT side, actually does not happen also on
the AdS side.

One may think that this argument suggests that the holographic dual
is nothing thermal in spite of the presence of the two boundaries as
opposed to the well-known example of AdS Schwarzschild black holes,
where the two boundaries are connected by an event horizon. However,
we showed that if we choose the ingoing boundary condition at the
other boundary, the holographic conductivity on a probe D3-brane
computed for the AdS boundary behaves like a metal at finite
temperature. Even though we are not completely sure if this choice
of boundary condition is physically allowed, this calculation
implies that the presence of the other boundary can make the
holographic dual theory thermal. On the other case, if we choose the
Dirichlet boundary instead, it is clear that the system behaves
like an insulator. Though our analysis of the entanglement entropy
implies that the latter choice corresponds to the correct choice,
we would like to leave this as a future problem.

Finally, since our background also includes an asymptotically flat
region, it may be interesting to reconsider our problem from the
viewpoint of holography for flat space, which has been recently
studied in \cite{LT} by employing the holographic entanglement
entropy.

\vspace{1cm}

\noindent {\bf Acknowledgments} We are grateful to Wei Li for
collaboration at an initial stage of this work and to Hong L\"{u} for
useful comments on the draft of this paper. TT would like to
thank Robert Myers for hospitality at Perimeter institute and useful
discussions, where part of this work was done. The work of MF is supported in part by JSPS Grant-in-Aid for Scientific Research No.22-1028. The work of TT is
also supported in part by JSPS Grant-in-Aid for Scientific Research
No.20740132, and by JSPS Grant-in-Aid for Creative Scientific
Research No.\,19GS0219. MF and TT is supported by World Premier
International Research Center Initiative (WPI Initiative), MEXT,
Japan.

\appendix

\section{Holographic Entanglement Entropy for a
geodesically traversable AdS Wormhole}

In the main text we analyzed the holographic entanglement entropy for
the AdS wormhole background (\ref{AWsol}) or equally (\ref{metric}).
As we showed, its behavior turns out to be different from our naive
expectation based on the quantum entanglement between two
holographic dual theories which live on the two boundaries.

The purpose of this appendix is to calculate the holographic
entanglement entropy for a geodesically traversable AdS wormhole.
We assume the following form of the metric \be
ds_0^2=(1+r^2)(-dt^2+dx^2+dy^2)+\dfrac{dr^2}{1+r^2}.\label{wma} \ee
Notice that the metric (\ref{wma}) is not a solution to
physically sensible gravity theories,\footnote{Indeed, we can easily find that
the metric (\ref{wma}) violates the null energy condition. This condition requires
$R_{\mu\nu}N^{\mu}N^{\nu}\geq 0$ for any null vector $N^{\mu}$. In particular
if we choose $N^t=1, N^r=1+r^2$ and $N^x=N^y=0$, we find $R_{\mu\nu}N^{\mu}N^{\nu}=-2$.}
but is just an example we take by hand.

 It is clear that (\ref{wma}) represents a wormhole spacetime.
 There are two asymptotically AdS boundaries at $r=\infty$ and
 $r=-\infty$. In this case, they are connected by time-like and null
 geodesics as opposed to the solution (\ref{metric}).

\begin{figure}[tb]
   \begin{center}
     \includegraphics[height=6cm]{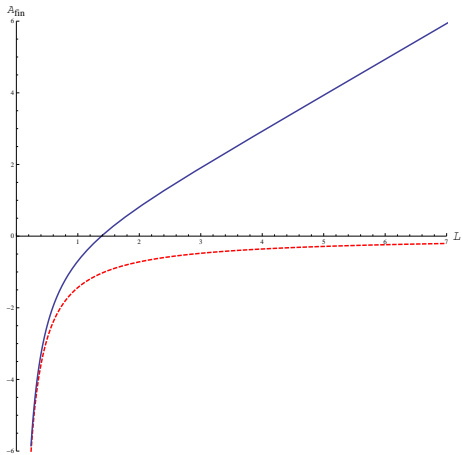}
   \end{center}
   \caption[.]{The $L$-dependence of the finite part of the area is plotted.
   The blue solid line shows the minimal area for the conventional wormhole \eqref{wma} while the red dashed line
   shows that for the pure AdS$_4$.}
\label{fig:3}
\end{figure}

We choose the subsystem $A$ to be the strip with width $L$ in the
$x$ direction, extending infinitely in $y$ direction. We want to
derive the $L$-dependence of the holographic entanglement entropy $S_A$.
The surface area in which $r$ is the function of $x$ is given by
\begin{align}
\text{Area}=2L_{\perp }\int^{L/2}_0 dx\sqrt{(1+r^2)^2+r^{\prime 2}},
\end{align}
where $L_{\perp }$ is the length of the surface in the
$y$-direction. Then we find the solution satisfies
\begin{align}
&r'=\dfrac{(1+r^2)\sqrt{(1+r^2)^2-H^2}}{H},
\end{align}
where $H$ is an integration constant. The turning point of the
minimal surface is given by $r=r_*$, which is related to $H$ as $H=1+r_*^2$.
Imposing $r(0)=r_*$ and $r(L/2)=\infty $, the length $L$ is given by
\begin{align}
L=2\int ^{\infty}_{r_*}dr \dfrac{1+r_*^2}{(1+r^2)\sqrt{(1+r^2)^2-(1+r_*^2)^2}} 
\end{align}
The area of the minimal surface is finally written
as
\begin{align}
{\rm Area}=2L_{\perp}\int_{r_*}^{1/\epsilon}dr \frac{1+r^2}{\sqrt{(1+r^2)^2-(1+r_*^2)^2}}
\end{align}
where $\ep$ is a UV cut-off.

We plot the finite part of the area as a function of $L$ in Fig.\ref{fig:3}. For
small $L$, the behavior is the same as that for the pure AdS$_4$.
However, for large $L$, the result shows that $S_A$ increases
linearly with $L$. This property looks very similar to the AdS black
holes rather than the AdS solitons. In other words, the holographic
dual of the wormhole background (\ref{wma}), if it exists, looks
like a thermal theory.

On the other hand, the AdS wormhole
solution (\ref{metric}), which has been discussed in the main part
of this paper, does not show this linear increase of $S_A$ and does
not look like thermal. Moreover, for large $L$, the entanglement
entropy is reduced compared with the pure AdS one.


\begin{thebibliography}{99}

\baselineskip=8pt






\bibitem{Maldacena}
  J.~M.~Maldacena,
  ``The large N limit of superconformal field theories and supergravity,''
  Adv.\ Theor.\ Math.\ Phys.\  {\bf 2} (1998) 231
  [Int.\ J.\ Theor.\ Phys.\  {\bf 38} (1999) 1113]
  [arXiv:hep-th/9711200];



\bibitem{GSWW}
  G.~J.~Galloway, K.~Schleich, D.~Witt and E.~Woolgar,
  ``The AdS / CFT correspondence conjecture and topological censorship,''
  Phys.\ Lett.\  B {\bf 505} (2001) 255
  [arXiv:hep-th/9912119].



\bibitem{Bergman:2008cv}
  A.~Bergman, H.~Lu, J.~Mei, C.~N.~Pope,
  ``AdS Wormholes,''
  Nucl.\ Phys.\  {\bf B810}, 300-315 (2009).
  [arXiv:0808.2481 [hep-th]].


\bibitem{Lu:2009ir}
  H.~Lu, J.~Mei and Z.~L.~Wang,
  ``GL(n,R) Wormholes and Waves in Diverse Dimensions,''
  Class.\ Quant.\ Grav.\  {\bf 26} (2009) 085020
  [arXiv:0901.0003 [hep-th]].



\bibitem{Wang:2009in}
  Z.~L.~Wang and H.~Lu,
  ``Most General Spherically Symmetric M2-branes and Type IIB Strings,''
  Phys.\ Rev.\  D {\bf 80} (2009) 066008
  [arXiv:0906.3439 [hep-th]].


\bibitem{WiYa}
  E.~Witten and S.~T.~Yau,
  ``Connectedness of the boundary in the AdS / CFT correspondence,''
  Adv.\ Theor.\ Math.\ Phys.\  {\bf 3} (1999) 1635
  [arXiv:hep-th/9910245].


\bibitem{MaMa}
  J.~M.~Maldacena and L.~Maoz,
  ``Wormholes in AdS,''
  JHEP {\bf 0402} (2004) 053
  [arXiv:hep-th/0401024].


\bibitem{AOP}
  N.~Arkani-Hamed, J.~Orgera and J.~Polchinski,
 ``Euclidean wormholes in string theory,''
  JHEP {\bf 0712} (2007) 018
  [arXiv:0705.2768 [hep-th]].


\bibitem{Dotti:2006cp}
  G.~Dotti, J.~Oliva and R.~Troncoso,
  ``Static wormhole solution for higher-dimensional gravity in vacuum,''
  Phys.\ Rev.\  D {\bf 75} (2007) 024002
  [arXiv:hep-th/0607062].


\bibitem{Ali:2009ky}
  M.~Ali, F.~Ruiz, C.~Saint-Victor and J.~F.~Vazquez-Poritz,
  ``Strings on AdS Wormholes,''
  Phys.\ Rev.\  D {\bf 80} (2009) 046002
  [arXiv:0905.4766 [hep-th]].

\bibitem{Arias:2010xg}
  R.~E.~Arias, M.~Botta-Cantcheff and G.~A.~Silva,
  ``Lorentzian AdS, Wormholes and Holography,''
  arXiv:1012.4478 [hep-th].

\bibitem{Skenderis:2009ju}
  K.~Skenderis and B.~C.~van Rees,
  ``Holography and wormholes in 2+1 dimensions,''
  Commun.\ Math.\ Phys.\  {\bf 301}, 583 (2011)
  [arXiv:0912.2090 [hep-th]].

\bibitem{MaE}
  J.~M.~Maldacena,
  ``Eternal black holes in anti-de Sitter,''
  JHEP {\bf 0304} (2003) 021
  [arXiv:hep-th/0106112].



\bibitem{RT}
  S.~Ryu and T.~Takayanagi,
  ``Holographic derivation of entanglement entropy from AdS/CFT,''
  Phys.\ Rev.\ Lett.\  {\bf 96} (2006) 181602
  [arXiv:hep-th/0603001];
``Aspects of holographic entanglement entropy,''
  JHEP {\bf 0608} (2006) 045
  [arXiv:hep-th/0605073].

\bibitem{Hubeny:2007xt}
  V.~E.~Hubeny, M.~Rangamani and T.~Takayanagi,
  ``A covariant holographic entanglement entropy proposal,''
  JHEP {\bf 0707}, 062 (2007)
  [arXiv:0705.0016 [hep-th]].


\bibitem{NRT}
  T.~Nishioka, S.~Ryu and T.~Takayanagi,
  ``Holographic Entanglement Entropy: An Overview,''
  J.\ Phys.\ A  {\bf 42} (2009) 504008
  [arXiv:0905.0932 [hep-th]].

\bibitem{CHM}
H.~Casini, M.~Huerta and R.~C.~Myers, ``Towards a derivation of
holographic entanglement entropy,'' arXiv:1102.0440 [hep-th].


\bibitem{EE}
L.~Bombelli, R.~K.~Koul, J.~H.~Lee and R.~D.~Sorkin,
  ``A Quantum Source of Entropy for Black Holes,''
  Phys.\ Rev.\ D {\bf 34}, 373 (1986);
M.~Srednicki,
  ``Entropy and area,''
  Phys.\ Rev.\ Lett.\  {\bf 71}, 666 (1993)
  [arXiv:hep-th/9303048].

\bibitem{Cardy}
C.~Holzhey, F.~Larsen and F.~Wilczek,
  ``Geometric and renormalized entropy in conformal field theory,''
  Nucl.\ Phys.\ B {\bf 424}, 443 (1994)
  [arXiv:hep-th/9403108];
P.~Calabrese and J.~L.~Cardy,
  ``Entanglement entropy and quantum field theory,''
  J.\ Stat.\ Mech.\  {\bf 0406}, P002 (2004)
  [arXiv:hep-th/0405152].

\bibitem{Review}
    P.~Calabrese and J.~Cardy,
  ``Entanglement entropy and conformal field theory,''
  J.\ Phys.\ A  {\bf 42} (2009) 504005
  [arXiv:0905.4013 [cond-mat.stat-mech]];
  H.~Casini and M.~Huerta,
  ``Entanglement entropy in free quantum field theory,''
  J.\ Phys.\ A  {\bf 42} (2009) 504007
  [arXiv:0905.2562 [hep-th]].

\bibitem{Sol}
  S.~N.~Solodukhin,
  ``Entanglement entropy of black holes,''
  arXiv:1104.3712 [hep-th].



\bibitem{Lu:2008zs}
  H.~Lu and J.~Mei,
  ``Ricci-Flat and Charged Wormholes in Five Dimensions,''
  Phys.\ Lett.\  B {\bf 666} (2008) 511
  [arXiv:0806.3111 [hep-th]].

\bibitem{NiTa}
  T.~Nishioka and T.~Takayanagi,
  ``AdS Bubbles, Entropy and Closed String Tachyons,''
  JHEP {\bf 0701} (2007) 090
  [arXiv:hep-th/0611035].

\bibitem{KKM}
  I.~R.~Klebanov, D.~Kutasov and A.~Murugan,
  ``Entanglement as a probe of confinement,''
  Nucl.\ Phys.\  B {\bf 796}, 274 (2008)
  [arXiv:0709.2140 [hep-th]].

\bibitem{PaPa}
  A.~Pakman and A.~Parnachev,
  ``Topological Entanglement Entropy and Holography,''
  JHEP {\bf 0807} (2008) 097
  [arXiv:0805.1891 [hep-th]].

\bibitem{Lata}
  A.~Velytsky,
  ``Entanglement entropy in d+1 SU(N) gauge theory,''
  Phys.\ Rev.\  D {\bf 77} (2008) 085021
  [arXiv:0801.4111 [hep-th]].

\bibitem{Latb}
  P.~V.~Buividovich and M.~I.~Polikarpov,
  ``Numerical study of entanglement entropy in SU(2) lattice gauge theory,''
  Nucl.\ Phys.\  B {\bf 802} (2008) 458
  [arXiv:0802.4247 [hep-lat]];
   ``Entanglement entropy in gauge theories and the holographic principle for
  electric strings,''
  Phys.\ Lett.\  B {\bf 670} (2008) 141
  [arXiv:0806.3376 [hep-th]].

\bibitem{Latc}
  Y.~Nakagawa, A.~Nakamura, S.~Motoki and V.~I.~Zakharov,
   ``Entanglement entropy of SU(3) Yang-Mills theory,''
  PoS {\bf LAT2009} (2009) 188
  [arXiv:0911.2596 [hep-lat]];
``Quantum entanglement in SU(3) lattice Yang-Mills theory at zero and finite
  temperatures,''
  PoS {\bf LATTICE2010} (2010) 281
  [arXiv:1104.1011 [hep-lat]].




\bibitem{Hartnoll:2008hs}
  S.~A.~Hartnoll and C.~P.~Herzog,
  ``Impure AdS/CFT correspondence,''
  Phys.\ Rev.\  D {\bf 77}, 106009 (2008)
  [arXiv:0801.1693 [hep-th]].
  
\bibitem{Fujita:2008rs}
  M.~Fujita, Y.~Hikida, S.~Ryu and T.~Takayanagi,
  ``Disordered Systems and the Replica Method in AdS/CFT,''
  JHEP {\bf 0812}, 065 (2008)
  [arXiv:0810.5394 [hep-th]].
  
\bibitem{KKY}
  S.~Kachru, A.~Karch and S.~Yaida,
  Phys.\ Rev.\  D {\bf 81} (2010) 026007
  [arXiv:0909.2639 [hep-th]].



\bibitem{JH}
  L.~Y.~Hung and Y.~Shang,
  ``On 1-loop diagrams in AdS space,''
  Phys.\ Rev.\  D {\bf 83} (2011) 024029
  [arXiv:1007.2653 [hep-th]].

\bibitem{AY}
A.~Adams and S.~Yaida,
  ``Disordered Holographic Systems I: Functional Renormalization,''
  arXiv:1102.2892 [hep-th].

\bibitem{RTU}
  S.~Ryu, T.~Takayanagi and T.~Ugajin,
  ``Holographic Conductivity in Disordered Systems,''
  arXiv:1103.6068 [hep-th].





\bibitem{Son:2002sd}
  D.~T.~Son and A.~O.~Starinets,
  ``Minkowski space correlators in AdS / CFT correspondence: Recipe and
  applications,''
  JHEP {\bf 0209}, 042 (2002)
  [arXiv:hep-th/0205051].

\bibitem{Policastro:2002se}
  G.~Policastro, D.~T.~Son and A.~O.~Starinets,
  ``From AdS / CFT correspondence to hydrodynamics,''
  JHEP {\bf 0209}, 043 (2002)
  [arXiv:hep-th/0205052].

\bibitem{KB}
A.~Karch and A.~O'Bannon, ``Metallic AdS/CFT,'' JHEP {\bf 0709}
(2007) 024  [arXiv:0705.3870 [hep-th]].


\bibitem{NiRT}
  T.~Nishioka, S.~Ryu and T.~Takayanagi,
  ``Holographic Superconductor/Insulator Transition at Zero Temperature,''
  JHEP {\bf 1003} (2010) 131
  [arXiv:0911.0962 [hep-th]].
\bibitem{Kobayashi:2006sb}
  S.~Kobayashi, D.~Mateos, S.~Matsuura, R.~C.~Myers and R.~M.~Thomson,
  ``Holographic phase transitions at finite baryon density,''
  JHEP {\bf 0702}, 016 (2007)
  [arXiv:hep-th/0611099].



\bibitem{LT}
  W.~Li and T.~Takayanagi,
  ``Holography and Entanglement in Flat Spacetime,''
  arXiv:1010.3700 [hep-th].





\end{thebibliography}
\end{document}